
\input phyzzx
\hoffset=0.2truein
\voffset=0.1truein
\hsize=6truein
\def\TITLEPAGE{\frontpagetrue}
\def\CALT#1{\hbox to\hsize{\tenpoint \baselineskip=12pt
        \hfil\vtop{
        \hbox{\strut CALT-68-#1}
        \hbox{\strut DOE RESEARCH AND}
        \hbox{\strut DEVELOPMENT REPORT}}}}

\def\CALTECH{
        \address{California Institute of Technology,
Pasadena, CA 91125}}
\def\TITLE#1{\vskip .5in \centerline{\fourteenpoint #1}}

\def\ABSTRACT#1{\vskip .2in \vfil \centerline{\twelvepoint
\bf Abstract}
        #1 \vfil}
\def\ENDTITLEPAGE{\vfil\eject\pageno=1}
\hfuzz=5pt
\tolerance=10000
\TITLEPAGE
\CALT{1752}
\TITLE{Topological Approach to Alice
Electrodynamics\footnote\dagger{This work supported in part
by the US Department of Energy under Contract No.
DE-AC03-81-ER40050}}

\vskip 15pt
\centerline{Martin Bucher, Hoi-Kwong Lo, and John Preskill}

\CALTECH
\ABSTRACT{We analyze the unlocalized ``Cheshire charge''
carried by ``Alice strings.''  The magnetic charge on a
string loop is carefully defined, and the transfer of
magnetic charge from a monopole to a string loop is analyzed
using global topological methods.  A semiclassical theory of
electric charge transfer is also described.}

\ENDTITLEPAGE
\eject

\def\C{{\cal C}}

\chapter{Introduction}

In a spontaneously broken non-abelian gauge theory, charge
conjugation can be a {\it local} symmetry.  That is, the
unbroken gauge group $H$ may contain both a $U(1)$ factor
generated by $Q$, and an element $X$ of the disconnected
component of $H$ such that $XQX^{-1}=-Q$.  Such a model
contains  topologically stable cosmic strings with a
remarkable property---when a charged particle is transported
around the string, the sign of its charge flips.  (The sign
of the charge is gauge dependent, but the feature that the
sign changes has an unambiguous and gauge--invariant
meaning.)  This string, which acts as a charge--conjugation
looking glass, was first discussed by A. S. Schwarz, who
dubbed it the ``Alice'' string.
\Ref\schwarz{A.S. Schwarz, ``Field Theories With No Local
Conservation of
Electric Charge," Nucl. Phys. {\bf B208}, 141 (1982); A.S.
Schwarz and Y.S. Tyupkin,
``Grand Unification and Mirror Particles,"
Nucl. Phys. {\bf B209}, 427 (1982).}
(The possibility that charge conjugation could be a local
symmetry was noted earlier by Kiskis.\Ref\kiskis{J. Kiskis,
``Disconnected Gauge Groups and the Global Violation of
Charge Conservation,''
Phys. Rev. {\bf D17}, 3196 (1978).})

A closed loop of Alice string can carry electric charge, and
the charge lost by a particle that winds around the string
is transferred to the loop.  A charged string loop is a
peculiar object.  It has a long--range electric field, from
which its charge can be inferred, yet there is no localized
source of charge anywhere on the string or in its vicinity.
\REFS\ginsparg{S. Coleman and P. Ginsparg, unpublished
(1982).}\REF\ABCMW{M. G. Alford, K. Benson, S. Coleman, J.
March-Russell, and F. Wilczek
``Interactions and Excitations of Non-Abelian Vortices,"
Phys.
Rev. Lett. {\bf 64}, 1632 (1990); {\bf 65}, 668
(E).}\REF\prekra{J. Preskill
and L. Krauss, ``Local Discrete Symmetry and
Quantum--Mechanical Hair,'' Nucl. Phys. {\bf B341}, 50
(1990).}\refsend
Such charge with no locally identifiable source has been
called ``Cheshire charge.''~~\refmark{\ABCMW}  An Alice
string can also carry magnetic Cheshire charge, and can
exchange magnetic charge with magnetic
monopoles.\REF\preskill{J. Preskill, ``Vortices and
Monopoles," in
{\it Architecture of the Fundamental
Interactions at Short Distances}, eds. P. Ramond and R. Stora
(North Holland, Amsterdam, 1987).}\REF\ABCMWSec{M. Alford,
K. Benson, S. Coleman, J. March-Russell, and F. Wilczek,
``Zero Modes of Non-Abelian Vortices,'' Nucl. Phys. {\bf
B349}, 414 (1991).}\refmark{\ginsparg,\preskill,\ABCMWSec}

The properties of Alice strings that carry Cheshire charge,
and the processes by which charge is exchanged between
strings and point particles, have been analyzed
previously.\refmark{\ABCMW-\ABCMWSec}.  In this analysis, it
is very convenient to employ the unitary gauge.  However, in
the presence of an Alice string, the gauge transformation
that imposes the unitary gauge condition is necessarily
singular; it introduces a gauge artifact surface on which
fields (the electric and magnetic fields in particular)
satisfy nontrivial boundary conditions.  At the price of
introducing this gauge--artifact singularity, one arrives at
an appealing and vivid description of the charge--transfer
phenomenon.

In this paper, we analyze Cheshire charge using a different
approach.  In the case of magnetic charge, we note that the
charge on a string is really a {\it topological} charge, and
that the transfer of charge from magnetic monopole to string
has an essentially topological origin.  The transfer of
topological charge can be described in a manifestly
gauge--invariant way.  By using global methods, one assuages
the
concern that the conclusions of previous work were an
unfortunate artifact of an illicit gauge choice.

Even in the case of electric charge, global methods provide
new insights.  We will trace the mechanism of electric
charge transfer to a generic topological property of
non-abelian vortices---namely, that when one vortex winds
around another, the quantum numbers of both are modified.

The rest of this paper is organized as follows:  In Section
2, we briefly review the simplest model that contains an
Alice string, and recall the analysis of Cheshire charge in
Ref.~\ABCMW-\ABCMWSec.  In Section 3, we
describe the long--range interactions between non-abelian
string loops, and use the properties of these interactions
to develop a semiclassical theory of Cheshire charge and
charge transfer.

In Section 4, we note the subtleties inherent in defining
magnetic charge in the presence of loops of Alice string.
For the purpose of defining the magnetic charge carried by a
particular string loop, it is convenient to introduce an
(arbitrary) ``basepoint,'' and a canonical surface (or
homotopy class of surfaces) that encloses the loop and is
tied to the basepoint.  In general, the canonical surface
can be chosen in topologically inequivalent ways, and the
enclosed magnetic charge depends on this choice.  It is just
this ambiguity that underlies the transfer of charge from a
magnetic monopole to a string loop.  We will find that, as a
monopole winds around a string loop, the canonical surfaces
that are used to define the magnetic charge of both the
monopole and the loop are deformed to new (topologically
inequivalent) surfaces.  Therefore, the charges defined by
the original canonical surfaces are modified; charge
transfer has taken place.

Section 5 contains some concluding remarks.

\chapter{Alice Strings}
The simplest model that contains an Alice string has gauge
group $SU(2)$ and a Higgs field $\Phi$ that transforms as
the 5-dimensional irreducible representation of $SU(2)$.  We
may express $\Phi$ as a real symmetric traceless $3\times 3$
matrix that transforms according to
$$
\Phi\to M\Phi M^{-1}~,~~~M\in SO(3)~.
\eqn\phitrans
$$
If $\Phi$ has an expectation value (in unitary gauge)
that can be
expressed as
$$
\left< \Phi \right> =v\cdot {\rm diag}~[1,\ 1,\ -2\ ]~,
\eqn\phivev
$$
then the unbroken subgroup of $SU(2)$ is $H=U(1)\times
_{S.D.}Z_2$.
The unbroken group $H$ has two connected components.  The
component connected to the identity
can be pictured as rotations about
a $z$-axis.  Since $SU(2)$ is a double cover of the rotation
group,
this component, which is isomorphic to $U(1),$
can be expressed as
$H_c=\{ \exp [i\theta Q]\  |~ 0\le \theta <4\pi \}$, where
$Q$ is the $SU(2)$ generator $Q={1\over 2}\sigma_3$.
There is also a connected component not connected to the
identity
of the form
$H_d=\{ X\exp [i\theta Q]\ |~ \ 0\le \theta <4\pi \} .$
This component consists of rotations by $180^\circ$ about
axes that lie
in the $xy$-plane.
($X$ is any such rotation.)  Each element $Y$ of $H_d$
anticommutes with $Q$, $YQY^{-1}=-Q$; it is
a ``charge--conjugation'' operator embedded
in the unbroken local
symmetry group.

The elements of $H_d$ represent the possible values of the
``magnetic flux''
of the topologically stable cosmic string excitations of the
theory
in 3+1 dimensions (or vortex excitations in 2+1 dimensions).
In general, the magnetic flux carried by a cosmic string
is an element of the unbroken group H that encodes the
result of parallel transport along a closed path that
encloses the string.  To define the magnetic flux we must
specify a
basepoint
$x_0$ and a closed loop
$C$ that starts and ends at $x_0$ and encircles the
string exactly once.\FIG\Caround{The curve $C$, starting and
ending at the point $x_0$,
encloses a loop of cosmic string.} (See Fig.~\Caround.)
Then the flux is given by the untraced Wilson loop
operator
$$
h(C,x_0)=P \exp\left(  i\int _{(C,x_0)}%
dx^i\ A_i \right)~.
\eqn\Wilop
$$
The flux takes values in $H(x_0)$, the subgroup of the
underlying group $G$
that stabilizes the condensate at the point $x_0$ (since
parallel transport around $C$ must return the condensate to
its original value).

One can determine what happens to the charge of a particle
that travels around an Alice string by considering the
behavior of the unbroken symmetry group $H(x_0)$ as it is
parallel transported around the string.\FIG\circle{A circle,
parametrized by $\phi$, encloses an Alice string.
Corresponding to each point of the circle is an unbroken
symmetry group $H(\phi)$ that stabilizes the condensate
$\Phi(\phi)$ at that point.}
Consider the situation depicted in Fig.~\circle, with a
single
Alice string enclosed by a  circle  parameterized by
$\phi , ~0\le \phi \le 2\pi .$
At each point on the circle labeled by $\phi ,$ there
is a subgroup $H(\phi )$ embedded in $G$
that stabilizes the condensate $\Phi(\phi )$ at that point.
The gauge vector potential $A_\mu $ relates these subgroups
through the equation
$$
H(\phi )= U(\phi ) H(0) U(\phi )^{-1}~,
\eqn\hphi$$
where
$$
U(\phi )=P\exp \left( i\int _0^\phi d\phi A_\phi \right) ~.
\eqn\uphi
$$
Note that $U(2\pi )=h(C,x_0).$
It is certainly true that $H(0)=U(2\pi )H(0)U(2\pi
)^{-1}$,
because $H(2\pi )=H(0),$ but the analogous relation does not
hold for
the generators of $H.$
Since $U(2\pi )\in H_d$, we have
$$
U(2\pi )\ Q\ U(2\pi )^{-1}=-Q~.
\eqn\Qflip
$$
An analogy can be made to the M\"obius strip to make it
apparent why $Q$ is deformed into $-Q$ upon parallel
transport around the circle. The
$U(1)$ subgroups $\{H(\phi )\}$ of $SO(3)$
can be represented as undirected lines in $\Re ^3$
through the origin that coincide with the axes of
the rotation of the $U(1)$ subgroups.
Choosing a generator $Q(\phi )$ for $H(\phi )$
at each $\phi $
is equivalent to choosing a direction for each of these
lines.
As $\phi $ varies from $0$ to $2\pi ,$
the lines are twisted into a M\"obius strip.
There is no continuous way to choose a direction
on each of them.

\def\M{{\cal M}}

The M\"obius twist in the unbroken symmetry group $H(x)$
described above may be discussed more formally in terms
of the ``global unrealizability" of the unbroken
symmetry.\REF\bala{A. Balachandran, F. Lizzi, and V.
Rodgers,
``Topological Symmetry Breakdown in Cholesterics,
Nematics, and
${}^3 He$," Phys. Rev. Lett. {\bf 52,} 1818
(1984).}\refmark{\bala,\ABCMW,\prekra}
Let ${\cal M}$ denote the spatial manifold consisting of
$\Re ^3$ with
the cores of the strings excised. At each point
$x\in {\cal M}$ is defined the unbroken symmetry group
$H(x)$ that stabilizes the Higgs condensate $\Phi(x).$
All these subgroups are isomorphic to the same abstract
group $H.$
This structure is a fiber bundle $E$ with model fiber $H$
over the base manifold $\M .$
The structure group of the bundle is also $H,$
and $H$ acts on the fibers by conjugation.
Locally, in any contractable
open subset $U\subset {\cal M},$ the fiber bundle
has the structure $U\times H.$ But generally
there does not exist a continuous mapping
$$
f: \M \times H \to E~.
\eqn\f
$$
This is because
the open sets $U_\alpha $ covering $\M $
can be patched together in a nontrivial way
using nontrivial transition functions.
In more
physical terms, a continuous mapping of the form $f$
is a ``global realization" of the unbroken symmetry $H$
considered as an abstract group. (In mathematical
language, such a mapping is known as a trivialization
of the fiber bundle $E.)$ Clearly, such a realization
is not possible in the presence of an Alice string,
because such a mapping $f$ would induce a continuous
choice of $Q(\phi )$ for $0\le \phi <2\pi $,
and we just showed that no such continuous choice exists.
(``Global unrealizability" of the unbroken symmetry also
occurs when there are monopoles with non-abelian magnetic
charge.
\Ref\color{P. Nelson and A. Manohar, ``Global Color is Not
Always
Defined," Phys. Rev. Lett. {\bf 50,} 943 (1983);
A. Balachandran, G. Marmo, N. Mukunda, J. Nilsson,
E. Sudarshan and F. Zaccaria, ``Monopole Topology and
the Problem of Color," Phys. Rev. Lett. {\bf 50,} 1553
(1983);
P. Nelson and S. Coleman, ``What Becomes of Global Color,"
Nucl. Phys. {\bf B237,} 1 (1984).})

The M\"obius twist implies that
a charged particle initially with charge $q$
will have charge $-q$ after winding around an Alice string.
Of course, the sign of the charge can be changed by a gauge
transformation, and therefore has no unambiguous physical meaning.
But the statement that the sign changes upon transport
around the string is gauge invariant and meaningful.
Suppose, for example, that two charges of like sign are
initially brought close together; they repel.
\FIG\chargearound{Initially two
particles carry charge of the same sign.  But after one of
the particles travels around the string, the particles carry
charge of opposite sign.} (See Fig.~\chargearound.)  Then
one charge travels around an Alice string while the other
stays behind.  When they are brought together again,
they attract.  Yet
the total charge, as measured by an observer far away from
the string loop and the point charges, cannot have changed.
Where did the missing charge go?

This puzzle is resolved by Cheshire
charge.\refmark{\ABCMW,\prekra}
In order to understand what happened to the charge, it is
convenient to choose a particular gauge---the unitary gauge
in which the Higgs field takes the value eq.~\phivev\
everywhere.  However, the gauge transformation that
implements the unitary gauge condition is singular; it has a
discontinuity, or cut,  on a surface that is bounded by the
string loop.
(In other words, one can transform to unitary gauge everywhere
outside a thin pancake that encloses the string loop.  Inside the
pancake, the Higgs field twists very rapidly,
and the gauge potential is very large.  The singularity
arises as the width of the pancake shrinks to zero.)
As a result, fields on the background of the
string loop obey peculiar boundary conditions---the
electromagnetic field changes sign on the cut, and charge of
a charged matter field flips there.

\FIG\cut{The surface $S$ is a cut at which the electric
field changes sign.  The loop in (b) carries Cheshire
charge.}
Because of the peculiar boundary conditions satisfied by the
electromagnetic field at the cut, there are solutions to the
classical field equations in which the cut appears to be a
source of electric (or magnetic) charge, as in Fig.~\cut.
There is not actually any measurable charge density on the
cut; the cut is an unphysical  gauge artifact.  Yet the
string loop is charged---it has a long range electric field
that can be detected by a distant observer.  This electric
field has no locally identifiable source; it is ``Cheshire
charge.''

\FIG\Cheshexch{A particle that initially has positive charge
travels through a loop of Alice string along the path shown
in (b).  The electric field during the process is indicated
schematically in (c)--(e).}
The charge transfer process is sketched in Fig.~\Cheshexch.
The initial electric field of a charge-$q$ particle in the
vicinity of a string loop is shown in Fig.~\Cheshexch a, and
Fig.~\Cheshexch c--e shows how the field changes as the
particle travels around the path in Fig.~\Cheshexch b.
When the particle crosses the cut, its apparent charge flips
from $q$ to $-q$, and the cut seems to acquire the
compensating charge $-2q$.  It is clear from the final
configuration in Fig.~\Cheshexch e that charge $2q$ has been
exchanged between the particle and the loop.

Yet there is no gauge--invariant way to pinpoint when the
charge transfer took place.  The configuration of the
electric field lines is gauge invariant, but the direction
of the arrows on the field lines is gauge dependent.  We can
move the cut by performing a singular gauge transformation; this
alters the apparent time of the charge transfer without
actually changing the physics of the process.

The charge transfer can be characterized in a
gauge--invariant manner, as follows:  The nontrivial
irreducible
representations of $H$ are two-dimensional, and can be
labeled by the absolute value of the $U(1)$ charge.  The
tensor product of two irreducible representations decomposes
into irreducible representations according to
$$
|q_1|\otimes |q_2|=|q_1+q_2|\oplus |q_1-q_2|~.
\eqn\decomp
$$
For the charge--loop system described above, the total
charge is $|q|$.  This charge determines (the absolute value
of) the electric flux through a large closed surface that
encloses the system, and is of course conserved during the
exchange process.  Initially, the loop is uncharged and the
particle has charge $|q|$.  The exchange process leaves
(the absolute value of)
the
charge of the particle unchanged, but produces an excitation
of the loop with charge $|2q|$.

So far, we have considered a particular model
with Alice strings. Much of the physics discussed in
this paper is independent of the details of that model. We
will
briefly describe
a more general class of models in which
Alice--like behavior occurs.\refmark{\schwarz}
Let the unbroken group $H$ to be a subgroup
of the simply-connected gauge group $G.$
Topologically stable cosmic strings occur only when $\pi
_0(H)$ is nontrivial,
so suppose that  $H$ has several connected components.
Groups of this sort may be constructed as the semi-direct
product of
a continuous part $H_c,$ which is a connected compact Lie
group,
and a discrete group $D.$
The semi-direct product $H_c\times _{S.D.}D$
is a generalization of a direct product,
defined by a group homomorphism
$$
\varphi :D\to {\rm Aut}[H_c]~,
\eqn\homomor
$$
where ${\rm Aut}[H_c]$ is the group of
automorphisms of $H_c.$
Group multiplication is defined using the rule
$$
(h_1,d_1)\circ (h_2,d_2)=(h_1\cdot \varphi
_{d_1}(h_2),d_1\cdot d_2)~.
\eqn\multrule
$$
Strings will have Alice properties if the mapping $\varphi $
is nontrivial.

In the example described earlier, $D=Z_2$ and
the nontrivial automorphism reverses the sign of the
generator $Q$ of $H_c=U(1)$.  As an example of generalized
Alice behavior, consider a model with
$$
H=\left[SU(2)_1\times SU(2)_2\right]\times_{S.D.} Z_2~,
\eqn\SUalice
$$
where the nontrivial automorphism is a ``parity'' operator
that interchanges  the two $SU(2)$ factors.  (With suitable
Higgs structure, the gauge group $G=SU(4)$ can be broken to
this $H$.)  This model contains an Alice-like string. If an
object with representation content $(R_1,R_2)$ under
$SU(2)_1\times SU(2)_2$ is transported around this string,
its representation content is changed to $(R_2,R_1)$, and
the missing quantum numbers are transferred to the string.

We should also note that a string might exhibit Alice-like
behavior, for dynamical reasons, even when such behavior is
not topologically required.\refmark{\ABCMW}  That is, the
flux of a dynamically stable string might assume a value $h$
that is not in the center of $H$, even though there are
elements of the center that lie in the same connected
component as $h$.  Then only the subgroup of $H$ that
commutes with the flux $h$ can be globally defined in the
presence of the string.  However, in this case, strictly
speaking, the position dependence of the unbroken symmetry
group $H(x)$ is not described by a topologically nontrivial
bundle.  This is because we can trivialize the
bundle by smoothly deforming the flux $h$
to an element of the center of $H$.
The bundle is nontrivial only if no element of the
center is contained in the same connected component as the
flux; that is, only if the Alice behavior is topologically
unavoidable.

\chapter{Electric Charge}
In this section, we describe the electrically charged Alice
string, and the charge transfer process, in semiclassical
language.\Ref\bucher{M. Bucher,
{\it On the Theory of Non-Abelian Vortices and Cosmic
Strings,} Ph.D. Thesis (Caltech, 1990).}

In quantum theory, the electric charge of a state reflects
the transformation properties of the state under global
gauge transformations.
The Alice string classical solution is not a charge
eigenstate, but it has a ``charge rotor'' zero mode.
Semiclassical quantization of the zero mode
is achieved by constructing linear combinations of the
classical string states that {\it do} have definite
charge.\refmark{\ABCMW,\prekra}
We need to worry, though, about what is meant by a
``global'' gauge transformation, since we have seen that
gauge transformations are not globally realizable.
Fortunately, for the purpose of defining the total charge of
a state, it is sufficient to consider a gauge transformation
that is constant on and outside a large sphere that encloses
all of the charged objects. Inside the sphere, we may deform
the gauge transformation so that it vanishes on the core of
each string, and on a surface bounded by each
string.\refmark{\prekra}  There is no topological
obstruction to constructing this gauge transformation.
Strictly speaking, since the flux of a string is defined
relative to a basepoint, we should think of the large sphere
not as a ``free'' surface, but rather as a surface tied to
the basepoint $x_0$.  That is, the gauge transformation
takes the same value at $x_0$ as on the sphere.  (If the
total {\it magnetic} charge enclosed by the sphere is
nonzero, then there is a further obstruction, so that the
gauge transformations in the disconnected component $H_d$
cannot be defined on the sphere.\refmark{\color}.  We defer
the discussion of magnetically charged string loops until
the next section, and suppose, for now, that the magnetic
charge is zero.)

The magnetic flux of the string, defined by eq.~\Wilop,
takes values in the disconnected component $H_d$ of the
unbroken group $H(x_0)$ that stabilizes the condensate at
the basepoint $x_0$.  In general, this flux transforms under
a transformation $g\in H(x_0)$ according to
$$
h(C,x_0)\to gh(C,x_0)g^{-1}~.
\eqn\gentrans
$$
In the case of an Alice string, let $\ket{\theta}$ denote
the string loop state with flux $h(C,x_0)=Xe^{i\theta Q}$.
Under a global $H$ transformation, the transformation
property eq.~\gentrans\ becomes
$$
U(e^{i\omega Q})\ket{\theta}=\ket{\theta-2\omega}~,
\eqn\statetrans
$$
$$
U(Xe^{i\omega Q})\ket{\theta}=\ket{2\omega-\theta}~,
\eqn\moretrans
$$
where $U$ is the unitary operator acting on Hilbert space
that represents the global gauge transformation.

One can construct linear combinations of these ``flux
eigenstate''
string states that transform irreducibly under $H$.
Let
$$
\ket{q}= \int_0^{4\pi} {d\theta\over\sqrt{4\pi}} ~
e^{i{\theta\over 2}q} \ket{\theta}
\eqn\cheshireD
$$
(where $Q$ is an integer). It transforms as
$$
U(e^{i\omega Q})\ket{q}=e^{i\omega q}\ket{q}~;
\eqn\cheshireE
$$
$$
U(Xe^{i\omega Q})\ket{q}= e^{i\omega q}\ket{-q}~.
\eqn\cheshireF
$$
The two states $\ket{q}$ and $\ket{-q}$ thus comprise the
basis for an
irreducible representation of $H$.

Only integer-$|q|$ representations of $H$ occur in this
decomposition; an Alice string cannot carry
half--odd--integer $|q|$.  String loops are invariant under
the center
of $SU(2)$, and so can have no ``two-ality.''

The semiclassical quantization of the charge rotor of the
Alice string is strongly reminiscent of the corresponding
treatment of bosonic superconducting strings.\Ref\witten{E.
Witten, ``Superconducting Strings,'' Nucl. Phys. {\bf B249},
557 (1985).}  But the physical properties of the string are
actually remarkably different.  Alice strings do not carry
persistent currents.  Instead, they carry electric charge
(or magnetic charge, as we will discuss in the next
section).

Now we will discuss the charge transfer process.  It will be
enlightening to imagine that the charged object that winds
through the string loop is itself a loop of Alice string.
Then the charge transfer can be regarded as a consequence of
a topological interaction between non-abelian string loops.
(We will see in the next section that magnetic charge
transfer results from a related topological interaction.)

\FIG\stringwind{The flux on the two string loops $\C_1$ and
$\C_2$ is defined with respect to the basepoint $x_0$ and
the paths $C_1$ and $C_2$.  When $\C_2$ winds through $\C_1$
as in (b), the paths $C_1'$ and $C_2'$ are dragged to $C_1$
and $C_2$.}
Consider the system of two string loops $\C_1$ and $\C_2$
shown in Fig.~\stringwind a.  Suppose that each string is a
flux eigenstate, with
$$
\eqalign{
&h(C_1,x_0)=h_1~,\cr
&h(C_2,x_0)=h_2~.\cr}
\eqn\twoflux
$$
Now suppose that the loop $\C_2$ winds through $\C_1$ as in
Fig.~\stringwind b.  To determine the magnetic flux of the
loops after the winding, it is convenient to consider the
paths $C_1'$ and $C_2'$ in Fig.~\stringwind c.  During the
winding procedure, these paths are dragged back to the paths
$C_1$ and $C_2$.
Therefore, the flux associated with the paths $C_1$ and
$C_2$ after the winding is the same as the flux associated
with the paths $C_1'$ and $C_2'$ before the winding.
One sees that
$C_1'=C_1$ and $C_2'={C_1}^{-1}\circ C_2\circ C_1$. (Our
convention is that $C_2\circ C_1$ denotes the path that is
obtained by traversing first $C_1$, then $C_2$.)  We therefore
find that, after
the winding, the flux carried by the string loops
is\REFS\wu{F. Wilczek and Y.-S. Wu, ``Space-Time Approach to
Holonomy Scattering,'' Phys. Rev.
Lett. {\bf 65}, 13 (1990).}\REF\martin{M. Bucher, ``The
Aharonov--Bohm Effect and Exotic Statistics for Non-Abelian
Vortices,'' Nucl. Phys. {\bf B350}, 163 (1991).}\REF\ALMP{M.
Alford, K.-M. Lee,
J. March-Russell, and J. Preskill, ``Quantum Field Theory of
Non-Abelian Strings and Vortices,'' Caltech preprint
CALT-68-1700 (1991).}\refsend
$$
\eqalign{
&h'(C_1,x_0)=h(C_1',x_0)=h_1~,\cr
&h'(C_2,x_0)=h(C_2',x_0)={h_1}^{-1} h_2 h_1~.\cr}
\eqn\fluxchange
$$

In the case of Alice strings, we denote by
$\ket{\theta_1,\theta_2}$ the two--string state with flux
$h_1=Xe^{i\theta_1 Q}$ and $h_2=Xe^{i\theta_2 Q}$.  Then, if
string 2 winds through string 1, eq.~\fluxchange\ becomes
$$
\ket{\theta_1,\theta_2}\to \ket{\theta_1,2\theta_1-
\theta_2}~.
\eqn\Alicechange
$$
If we construct charge eigenstates as in eq.~\cheshireD, we
find from eq.~\Alicechange\ that the effect of the winding
is
$$
\ket{\theta_1,q_2}\to e^{i\theta_1 q_2}\ket{\theta_1, -
q_2}~,
\eqn\thetaq
$$
and
$$
\ket{q_1,q_2}\to\ket{q_1+2q_2,-q_2}~.
\eqn\chargechange
$$
Just as in the classical analysis of Section 2, the sign of
$q_2$ flips, and loop 1 acquires a compensating charge.

Of course, we can also analyze (somewhat more
straightforwardly) the case in which the charge that winds
is a point charge rather than a charged loop.  Then
eq.~\thetaq\ follows directly from the gauge transformation
property of the charged particle.

\chapter{Magnetic Charge}
\section{Twisted Flux}
In the above discussion of semiclassical quantization, we
assumed that the magnetic flux was a constant along the
string.  But if the unbroken group $H$ is continuous, as in
the Alice case, the flux can vary as a function of position
along the string loop.  Furthermore, if $H$ is not simply
connected,
then the flux might trace out a noncontractible closed path in
$H$.  Then the string loop evidently carries a type of
topological charge.  This charge is precisely the magnetic
charge of the loop.

To define this charge carefully, we should, as usual, select
an arbitrary basepoint $x_0$ and consider the magnetic flux
defined by eq.~\Wilop.  As the path $C$ is smoothly
deformed
with the basepoint $x_0$ held fixed, this flux varies
smoothly in a given connected component of the group
$H(x_0)$.\FIG\degtorus{The family of closed paths
$\{ C_\phi \ | \ 0\le\phi <2\pi \} $
sweeps out a degenerate torus that encloses the Alice string
loop.}  To be specific, consider the family of paths
$\{C_\phi~|~ 0\le\phi<2\pi\}$, shown in Fig.~\degtorus.  These
paths sweep out a degenerate torus that encloses the string
loop.  This family $\{C_\phi\}$ is associated with a closed
path in $H(x_0)$ that begins and ends at the identity;
namely,
$$
h(C_\phi,x_0)h^{-1}(C_{\phi=0},x_0)~,~~~0\le\phi<2\pi~.
\eqn\grouploop
$$
We have thus found a natural way of mapping a two-sphere
that encloses the string loop to a closed loop in $H_c$, the
component of $H$ connected to the identity.

\FIG\deffam{A family of closed paths $\{C'_\phi\}$ obtained
by smoothly deforming the family $\{C_\phi\}$.}By smoothly
deforming the family $\{C_\phi\}$, we may obtain the family
of closed paths $\{C'_\phi\}$ shown in Fig.~\deffam.
Loosely
speaking, $h(C'_\phi,x_0)$ is the flux carried by the string
at the point where $C'_\phi$ wraps around the core of the
string.  Thus we see that the homotopy class of the path
defined by
eq.~\grouploop\ describes how the flux of the string twists
as a function of position along the string.

\FIG\lubkin{A family of loops $C^{\prime \prime }_\phi $
that sweeps over the surface of a sphere. The
loops
$C^{\prime \prime }_0$ and $C^{\prime \prime }_{2\pi }$ are
degenerate.}  On the other hand, the family $\{C_\phi\}$ is
equivalent to the family of paths $\{C''_\phi\}$ shown in
Fig.~\lubkin.  But this is just the family of paths used by
Lubkin\REF\lubkin{E. Lubkin, ``Geometric Definition of Gauge
Invariance,''
Ann. Phys. {\bf 23,} 233 (1963).}\REF\coleman{S.
Coleman,``The Magnetic Monopole Fifty Years Later,"
in {\it The Unity of the Fundamental Interactions}, ed. A.
Zichichi
(Plenum, New York,
1983).}\refmark{\lubkin,\coleman,\preskill} to define the
topological $H_c$ magnetic flux inside a two-sphere.  We
learn that the element of $\pi_1(H_c)$ that characterizes
how the magnetic flux of the string twists is the same as
the magnetic charge on the loop.\refmark{\bucher,\preskill}

More generally, in the presence of many string loops and
pointlike monopoles, we can define the magnetic charge
inside any region $R$ whose boundary $\partial R$ is
homeomorphic to $S^2$.  The result is a homomorphism
$$
h^{(2)}:~\pi _2[\M , x_0] \to \pi _1[H_c(x_0)]~,
\eqn\mapping
$$
where $\M $ denotes the manifold that is obtained when all
string loops and monopoles are removed from $\Re^3$.

\section{Role of the Basepoint}
We should now explain why it is important to specify a
basepoint $x_0$ for the purpose of defining the magnetic
charge.  Naively, it seems that it should be possible to
define the magnetic charge enclosed by a ``free'' surface
that is not tied to any basepoint, since the enclosed charge
is just the magnetic flux through the surface.  But trouble
arises if we allow the magnetic charges to move.  We can
deform the free surface so that it is never crossed by any
moving magnetic monopoles or string loops.  Nevertheless,
the magnetic flux through the surface can change if the
surface winds through an Alice string loop.

It will be easier to keep track of charge transfer processes
if we define magnetic charge using a surface that is tied to
a basepoint.  As the charges move, we can again deform the
surface so that no monopoles or strings cross it, while
keeping the basepoint fixed (as long as no monopoles or
strings cross the basepoint).  Then the magnetic charge
enclosed by the surface remains invariant.  However, when a
monopole winds around a string loop, the surface enclosing
the monopole becomes deformed to a new, topologically
inequivalent surface.   We can then find how the charge of
the monopole has changed by expressing the new surface in
terms of the old one.  This procedure is closely analogous
to our discussion in Section 3 of how the flux of a loop is
modified when it winds around another string.  There we
defined the flux using a standard {\it path} that became
deformed to a new path due to the winding.  We can analyze
the exchange of magnetic charge using a similar strategy,
except that a {\it surface}, rather than a path, is used to
define the charge.

\FIG\surfaces{The free surface $\bar\Sigma$ in (a) can be
threaded to the basepoint $x_0$ in inequivalent ways, two of
which are illustrated in (b) and (c).  The surface (c) can
be deformed to (d), which differs from (b) by the degenerate
tube $\beta$ that begins and ends at the basepoint.}
In order to define the magnetic charge enclosed by a free
surface $\bar \Sigma$ that is homeomorphic to $S^2$, then,
we specify not just the surface, but also a path that
attaches the surface to the basepoint $x_0$.
Of course, this path can be chosen in many topologically
inequivalent ways; the different choices are classified by
$\pi_1[M,x_0]$.  Thus, $\pi_1[M,x_0]$ classifies the
ambiguity in associating a free surface with an element of
$\pi_2[M,x_0]$.  There is a corresponding ambiguity in the
value of the magnetic charge (given by the homomorphism
$h^{(2)}$ defined in eq.~\mapping) that is associated with a
free surface.  We  resolve this ambiguity by simply choosing
a standard convention for the path from the free surface to
the basepoint, and sticking with this convention throughout
the process under study.

The ambiguity is illustrated by Fig.~\surfaces, which shows
two inequivalent surfaces $\Sigma$ and $\Sigma'$ with basepoint
$x_0$ that are
obtained by ``threading'' the free surface $\bar \Sigma$
to the basepoint in two different ways.
As shown in Fig.~\surfaces d,
the surface $\Sigma'$  can be deformed to a
degenerate tube, beginning and ending at $x_0$, joined to the
surface $\Sigma$.  Since the degenerate
tube is equivalent to a closed path $\beta$, we may say that
the two surfaces differ by an element of $\pi_1[M,x_0]$.

The ambiguity in associating a free surface with an element
of $\pi_2[M,x_0]$ can be characterized by a natural
homomorphism
$$
\tau :\pi _1[\M ,x_0] \to %
{\rm Aut}\left( \pi _2[\M , x_0]\right)
\eqn\definetau
$$
that takes (homotopy classes of) closed paths to
automorphisms of $\pi_2[M,x_0]$.
The mapping $\tau$ is defined in the following way:  Let
$\beta\in\pi_1[M,x_0]$ and $\Sigma\in \pi_2[M,x_0]$.
(Below we use the symbols $\beta$ and $\Sigma$ to denote
both homotopy equivalence classes and particular
representatives of the classes.)  Then $\tau_{\beta}$ is an
automorphism that takes $\Sigma$ to a new surface $\Sigma'$,
$$
\tau_{\beta}:~\Sigma\to\Sigma'~,
\eqn\taubeta
$$
where $\Sigma'$ is the surface $\Sigma$ with the degenerate
tube $\beta$ added on.  More precisely, let $\beta(t),~0\le
t\le 1$ be a parametrized path, with
$\beta(0)=\beta(1)=x_0$, and let $\Sigma(\theta,\phi),~0\le
\theta\le\pi,~0\le\phi\le 2\pi$ be a parametrized surface
with $\Sigma(0,\phi)=x_0$.  Then the new surface $\Sigma'$
is
$$
 \Sigma ' (\theta ,\phi )=\cases { %
\beta (2\theta /\pi ) &if $0\le \theta \le \pi /2~,$ \cr
\Sigma (2\theta -\pi , \phi )%
&if $\pi /2 \le \theta \le \pi ~.$ \cr }
\eqn\newsigma
$$

Now consider how changing the threading of a free surface to
the basepoint modifies the magnetic charge enclosed by the
surface.  Recall that eq.~\Wilop\ maps a path that begins
and ends at the basepoint to an element of the group
$H(x_0)$.  If the path is deformed to a homotopically
equivalent path, the group element remains in the same
connected component of the group.  Thus, eq.~\Wilop\ defines
a homomorphism
$$
h^{(1)}:~\pi_1[M,x_0]\to\pi_0[H(x_0)]~.
\eqn\pathhomo
$$
If the surface $\Sigma$ is changed to the surface $\Sigma'$
by adding the degenerate tube $\beta$, then the magnetic
charge enclosed by the new surface is related to the
magnetic charge enclosed by the original surface according
to
$$
h^{(2)}(\Sigma')=
h^{(1)}(\beta)^{-1}~h^{(2)}(\Sigma)~h^{(1)}(\beta)~.
\eqn\pathmagchange
$$
That is, $h^{(2)}(\Sigma')$ is the closed path in $H_c$
(beginning and ending at the identity) that is obtained
when $h^{1}(\beta)$ acts on the closed path $h^{(2)}(\Sigma)$ by conjugation.
In the case of the Alice string, eq.~\pathmagchange\ simply
says that, if $\beta$ is a path that winds around a string
loop, then the magnetic charges enclosed by $\Sigma$ and
$\Sigma'$ differ by a sign.

\section{Charge Transfer}
Eq.~\pathmagchange\ is the key to understanding the magnetic
charge transfer process, as we will  show.  First, though,
we should recall that $\pi_2[M,x_0]$  has a group structure
that allows
magnetic charge to be added.
The group multiplication law,
$$
\circ : \pi _2[\M ,x_0] \times \pi _2[\M ,x_0]\to \pi _2[\M,
x_0]~,
\eqn\mult
$$
can be defined as
$$
 \bigl( \Sigma _1\circ \Sigma _2\bigr) (\theta ,\phi
)=\cases { %
\Sigma _1(2\theta, \phi) &if $0\le \theta \le \pi /2~,$
\cr
\Sigma _2(2\theta -\pi , \phi )%
&if $\pi /2 \le \theta \le \pi ~.$ \cr }
\eqn\repmult$$
where $\Sigma _1 ,$ $\Sigma _2$, and
$ \Sigma _1 \circ \Sigma _2$
are homotopy equivalence class representatives.
Group multiplication in $\pi _2$ is commutative.
Group inversion may be expressed in terms of class
representatives as
$$ \Sigma ^{-1}(\theta ,\phi )=
\Sigma (\pi -\theta ,\phi )~.
\eqn\inverse
$$

\FIG\magfig{The magnetic flux of the string loops $\C_1$ and
$\C_2$ is defined in terms of the paths $\beta_1$ and
$\beta_2$ shown in (a), and the magnetic charges of the
loops are defined in terms of the surfaces $a_1$ and $a_2$;
the paths and the surfaces are based at the point $x_0$.
When $\C_2$ winds through $\C_1$ as in (b), the surface
$a_1'$ shown in (c) is dragged to $a_1$, and the surface
$a_2'$ shown in (d) is dragged to $a_2$.  The arrows on the
surfaces indicate outward--pointing normals.}
We turn to  the situation depicted in Fig.~\magfig.  Two
string loops $\C_1$ and $\C_2$ are shown.  We denote by
$\beta_1$ and $\beta_2$ two standard paths, beginning
and ending at the basepoint $x_0$, that wind around the
string loops.  (These are elements of $\pi_1[M,x_0]$.)  We
denote by $a_1$ and $a_2$ two standard surfaces, based
at $x_0$, that enclose the string loops.  (These are
elements of $\pi_2[M,x_0]$.)
The magnetic charges of the two loops, given by the
homomorphism eq.~\mapping, are $h^{(2)}(a_1)$ and
$h^{(2)}(a_2)$, respectively.

Now suppose that the loop $\C_2$ winds through the loop
$\C_1$ along the path shown in Fig.~\magfig b.  We want to
determine the magnetic charges of the two loops after this
winding.  To do so, consider the surfaces $a_1'$ and $a_2'$
shown in Fig.~\magfig c-d.  During the winding, these
surfaces are dragged back to the surfaces $a_1$ and $a_2$,
if the surfaces are deformed so that no surface ever touches
a string loop.  Therefore, the magnetic charge enclosed by
$a_1$, after the winding, is the same as the magnetic charge
enclosed by $a_1'$, before the winding.  Similarly, the
magnetic charge enclosed by $a_2$, after the winding, is the
same as the magnetic charge enclosed by $a_2'$, before the
winding.

\FIG\magfigdef{Deformations of the surfaces shown in
Fig.~\magfig c-d.  In (a), the surface $a_2'$ has been
deformed to the degenerate tube $\beta_1$ plus the surface
$a_2$.  In (b), the surface $a_1'$ has been deformed to the
surface $a_1\circ a_2$ that encloses both loops, plus the
inverse of $a_2'$ (that is, $a_2'$ with the orientation
reversed); the surface $(a_2')^{-1}$ is the sum of the
degenerate tube $(\beta_1)^{-1}$ and the surface $(a_2)^{-
1}$.}
It remains to find the magnetic charges enclosed by $a_1'$
and $a_2'$ before the winding.  Fig.~\magfigdef a shows a
deformation of $a_2'$ that makes it manifest that $a_2'$ can
be expressed as
$$
a_2'=\tau_{\beta_1}(a_2)~,
\eqn\newsurf
$$
where $\tau_{\beta_1}$ is the automorphism of $\pi_2[M,x_0]$
defined by eq.\taubeta --\newsigma.  In Fig. \magfigdef b,
the surface $a_1'$ is expressed as the sum of two surfaces.
The first (outer) surface is just $a_1\circ a_2$, the
surface that encloses both loops.  The second (inner)
surface is $(a_2')^{-1}$; it is the same as $a_2'$, except
with the opposite orientation.  We see that
$$
a_1'=a_1\circ a_2\circ (a_2')^{-1}~.
\eqn\othernew
$$
Finally, we apply eq.~\pathmagchange\ to find the magnetic
charges after the winding; the result is
$$
\eqalign{
&{h^{(2)}}'(a_1)=h^{(2)}(a_1')=h^{(2)}(a_1\circ a_2)
\Bigl[{h^{(2)}}'(a_2)\Bigr]^{-1}~,\cr
&{h^{(2)}}'(a_2)=h^{(2)}(a_2')=
h^{(1)}(\beta_1)^{-1}~h^{(2)}(a_2)~h^{(1)}(\beta_1)~.\cr}
\eqn\magchaexch
$$
Of course, the total magnetic charge is unchanged, because
$h^{(2)}(a_1\circ a_2)={h^{(2)}}'(a_1\circ a_2)$.

In the case of the Alice string, the magnetic charge can be
labeled by an integer $p$---the charge in units of the Dirac
charge.  If the magnetic charges on the string loops are
initially $p_1$ and $p_2$, and then loop 2 winds through
loop 1, eq.~\magchaexch\ says that the charges become
modified according to
$$
\ket{p_1,p_2}\to \ket{p_1+2p_2,-p_2}~,
\eqn\alicemagex
$$
in accord with the analysis in Section 2.

\section{Dyons}
We may also consider dyonic Alice string loops, that carry
both magnetic and electric charge.  The classical
magnetically charged Alice string loop has a charge rotor
zero mode, just like the magnetically neutral loop
considered in Section 3, and we may proceed with
semiclassical quantization in the same manner as before.
The only difference from the previous discussion is that,
for the magnetically charged loop, there is a topological
obstruction to defining global gauge transformations in the
disconnected component of the unbroken group $H$, similar to
the obstruction discussed in Ref.~\color. (The obstruction
occurs because the automorphism that reverses the sign of
$Q$ is incompatible the matching condition of a magnetic
monopole.) Thus, we obtain
states that transform irreducibly under the connected
component $H_c=U(1)$, but the states
do not transform as representations of the full
group.

By decomposing the classical  string with magnetic charge
$p$ into irreducible representations of $U(1)$, as in
Section 3, we find states $\ket{q,p}$ with electric charge
$q$, where $q$ is any integer.  Reanalyzing the charge
transfer process, we find that, when loop 2 winds through
loop 1, the charge assignments change according to
$$
\ket{q_1,p_1;q_2,p_2}\to\ket{q_1+2q_2,p_1+2p_2;-q_2,-p_2}~.
\eqn\dyonexch
$$
Naturally, both magnetic charge and electric charge are
exchanged.

We will comment briefly on how the analysis is modified
when the vacuum $\theta$-angle is nonzero.  The nonvanishing
vacuum angle alters the $U(1)$ transformation properties of
states with nonzero magnetic charge, so that eq.~\cheshireE\
is replaced by\Ref\witten{E. Witten, ``Dyons of Charge
$e\theta/2\pi$,'' Phys. Lett. {\bf 86B},
283 (1979); F. Wilczek, ``Remarks on Dyons,''
Phys. Rev. Lett. {\bf 48}, 1146
(1982).}
$$
U(e^{i\omega Q})\ket{q,p}=\exp\left[i\omega
\left(q+{\theta\over 2\pi}p\right)\right]\ket{q,p}~,
\eqn\thetalaw
$$
where $Q$ is the $U(1)$ generator, and $q$ is the charge of
the state defined in terms of the electric flux through the
surface at spatial infinity.  Thus, for magnetically charged
string loops, as for all magnetically charged objects, the
charge spectrum is displaced away from the integers by $-
\theta p/2\pi$.
But otherwise, the discussion of electric and magnetic
charge transfer is not altered; in particular,
eq.~\dyonexch\
still applies.

\section{Linked Loops}
The homomorphism defined in eq.~\mapping\ assigns a magnetic
charge to
any region whose boundary is homeomorphic to $S^2$.  But if
two
string loops link, the magnetic charge on each
individual loop is not well defined in general.  Only the
total magnetic charge of the two loops can be defined.
The magnetic field of a pair of linked loops has some
interesting properties that we will briefly discuss.

In general, two non-abelian string loops can link only if
the commutator of their fluxes is in the connected component
of the unbroken group.\foot{We thank Tom Imbo
for a helpful discussion about this.}  This feature is a
consequence of the ``entanglement'' phenomenon.
Suppose that a string loop with flux $h_1$ and a string loop
with flux $h_2$ cross each other, and become linked.  After
they cross, a flux $h_1h_2{h_1}^{-1}{h_2}^{-1}$ must flow from
one loop to the other.\REF\toulouse{V. Po\'enaru and G.
Toulouse, ``The Crossing of Defects in Ordered Media and the
Topology of 3-Manifolds,'' J. Phys. (Paris) {\bf 38}, 887
(1977).}\REF\mermin{N.D. Mermin, ``The Topological Theory of
Defects in
Ordered Media," Rev. Mod. Phys. {\bf 51,} 591
(1979).}\refmark{\toulouse,\mermin,\ALMP}  If this
commutator is not in the connected component of $H$, then
the commutator flux is itself confined to a stable string.
Thus, the two loops must be connected by a segment of string
that carries the commutator flux.
On the other hand, if the commutator is in the connected
component of $H$, then the commutator flux is unconfined,
and the flux will spread out uniformly over the $h_1$ and
$h_2$ loops.  The linked loops will have a long--range
magnetic dipole field, though the total magnetic charge of
the linked pair is zero.

In the case of the Alice string, consider two linked loops
that carry flux $Xe^{i\theta_1 Q}$ and $Xe^{i\theta_2 Q}$,
respectively.  The commutator flux $e^{i2(\theta_2-\theta_1)
Q}$ is in $H_c$, so that linking is allowed.  The strength of
the dipole field is proportional to $\theta_2-\theta_1~({\rm
mod}~ 2\pi)$.  If we fix the positions of the loops and
specify initial values for $\theta_1$ and $\theta_2$, then,
since the dipole field costs magnetostatic energy, the angle
$\theta_2-\theta_1$ will oscillate and the dipole field
will become time dependent.  These oscillations will cause
emission of radiation, and $\theta_2-\theta_1$ will decay,
eventually approaching zero.

\chapter{Concluding Remarks}
In any model in which a connected gauge group $G$ breaks to
a group $H$ that has a disconnected component, there will be
topologically stable strings.  If, in addition, $H$ contains
noncontractible closed paths, then the magnetic flux of a
string loop can have a topologically stable twist.  Thus,
the string loop can carry magnetic charge.  Note, in
particular, that in order for a string loop to be capable of
carrying magnetic charge,
there is no need for charge conjugation to be a local
symmetry.

But if the string is not an Alice string, the magnetic
charge will not be Cheshire charge---instead, the magnetic
charge will be localized on the string.  A loop with
Cheshire magnetic charge will have Coulomb energy of order
$$
E_{\rm Coulomb}\sim {p^2\over e^2~R}~,
\eqn\coulomb
$$
where $p$ is the charge in units of the Dirac magnetic
charge, $e$ is the gauge coupling, and $R$ is the size of
the loop.  If the charge is localized on the string, the
Coulomb energy is enhanced by a factor of order ${\rm
log}(R/r)$, where $r$ is the thickness of the string.

When the charge $p$ is large,  classically stable string
loop configurations can be constructed, such that the
Coulomb potential energy prevents the loop from collapsing.
If the charge is Cheshire charge, then the size $R$ and mass
$m$ of a stable loop are, in order of magnitude,
$$
R\sim \left({p\over e}\right)\kappa^{-{1\over 2}}~,
{}~~~m\sim \left({p\over e}\right)\kappa^{1\over 2}~,
\eqn\stable
$$
where $\kappa$ is the string tension.  Though classically
stable, these string loops are not expected to be absolutely
stable; they will emit elementary monopoles via a quantum
tunneling process, assuming that the emission is
kinematically allowed.

As we have seen, any phase transition that produces Alice
strings must also produce magnetic monopoles.  This
observation significantly restricts the role that Alice
strings can play in cosmology.  The process that produces
the strings will also produce an unacceptably large
abundance of monopoles.\Ref\monopreskill{J. Preskill,
``Cosmological Production of Superheavy Magnetic Monopoles,"
Phys. Rev. Lett. {\bf 43,} 1365 (1979).}  If such a process
occurred in the very early universe, it (presumably) must
have been followed by inflation that reduced the monopole
abundance to an acceptable level.  But, of course, the
inflation would also make Alice strings extremely scarce.

One caveat should be mentioned.  The remark in the previous
paragraph applies to any model such that the unbroken gauge
group $H$  contains a $U(1)$ factor and a charge conjugation
operator that reverses the sign of the $U(1)$ generator $Q$.
But it need not apply to models that exhibit the generalized
Alice-like behavior considered at the end of Section 2.  In
particular, a model with the unbroken group $H=\left[SU(2)_1\times
SU(2)_2\right]\times_{S.D.} Z_2$ contains generalized Alice
strings.  But since $H$ is simply connected, this model
contains no magnetic monopoles.

Finally, we remark that the discussion of magnetic charge
transfer in Section 4 also applies to the line and point
defects that arise when a {\it global} symmetry group $G$
becomes spontaneously broken to a subgroup $H$.  (Such
defects can occur in certain condensed matter
systems, such as nematic liquid crystals.\refmark{\mermin})
By a standard argument,\refmark{\coleman,\preskill,\mermin}
the magnetic charge, classified by $\pi_1[H]$, is seen to be
equivalent to the topological charge of the order parameter
$\Phi$, classified by $\pi_2[G/H]$ (assuming that $G$ is
simply connected).  Thus, our previous analysis applies,
without modification, to the transfer of topological charge
between a ``global monopole'' and a ``global Alice string.''

Recently, Brekke, Fischler, and Imbo\Ref\imbo{L. Brekke, W.
Fischler, and T. Imbo, ``Alice Strings,
Magnetic Monopoles, and Charge Quantization," Harvard
Preprint
HUTP-91/A042 (1991).} have independently investigated the
properties of magnetically charged Alice strings.

\bigskip
We thank Mark Alford, Rick Davis, Tom Imbo, Kai-Ming Lee,
John March-Russell, Sandip Trivedi,
and Piljin Yi for interesting
discussions.

\refout
\figout
\bye